\documentclass[pra,10pt,reprint,superscriptaddress,showpacs]{revtex4-1}

\usepackage{epsfig}
\usepackage{amsmath}
\usepackage{graphicx}
\usepackage{graphics}
\usepackage{float}
\usepackage{amssymb}
\usepackage[usenames,dvipsnames]{color}
\usepackage{natbib}
\usepackage[outdir=./]{epstopdf}
\usepackage{wrapfig}
\usepackage{subfigure}
\usepackage{bbold}
\usepackage{xcolor}
\usepackage[makeroom]{cancel}
\usepackage[utf8]{inputenc}
\usepackage{mathtools}
\usepackage{dsfont}
\usepackage{verbatim}

\newcommand{\bra}[1]{\left\langle #1\right|}
\newcommand{\ket}[1]{\left| #1\right\rangle}

\newcommand{\beq}{\begin{equation}}
\newcommand{\eeq}{\end{equation}}

\newcommand{\tr}{\text{Tr}}

\newcommand{\subfigimg}[3][,]{%
  \setbox1=\hbox{\includegraphics[#1]{#3}}
  \leavevmode\rlap{\usebox1}
  \rlap{\hspace*{200pt}\raisebox{\dimexpr\ht1-5\baselineskip}{#2}}
  \phantom{\usebox1}
}

\begin{document}

\title{The quantum Zeno and anti-Zeno effects: from weak to strong system-environment coupling}

\author{Bilal Khalid}
\email{m.bilalkhalid39@gmail.com}
\affiliation{School of Science \& Engineering, Lahore University of Management Sciences (LUMS), Opposite Sector U, D.H.A, Lahore 54792, Pakistan}

\author{Adam Zaman Chaudhry}
\email{adam.zaman@lums.edu.pk}
\affiliation{School of Science \& Engineering, Lahore University of Management Sciences (LUMS), Opposite Sector U, D.H.A, Lahore 54792, Pakistan}

\begin{abstract}

By repeatedly measuring a quantum system, the evolution of the system
can be slowed down (the quantum Zeno effect) or sped up (quantum anti-Zeno effect). We study these effects for a single two-level system coupled to a collection of harmonic oscillators. Previously, such systems have been studied in both the weak and the strong system-environment coupling regimes. In this paper, we apply a polaron transformation in a manner that allows us to study the quantum Zeno and anti-Zeno effects for a large variety of system-environment parameters. Using this approach, we reproduce previous results for the weak and strong system-environment coupling regimes. Moreover, as long as the environment is super-Ohmic, we show how our approach can be used to explore regimes such as the moderate system-environment coupling regime that could not be investigated before in a straightforward manner.

\end{abstract}

\date{\today}

\maketitle

\section{Introduction}

The quantum Zeno effect (QZE) is the slowing down of the temporal evolution of a quantum system when subjected to very frequent measurements ~\cite{Sudarshan1977,FacchiPhysLettA2000,FacchiPRL2002, FacchiJPA2008, WangPRA2008, ManiscalcoPRL2008, FacchiJPA2010, MilitelloPRA2011, RaimondPRA2012, SmerziPRL2012, WangPRL2013, McCuskerPRL2013, StannigelPRL2014, ZhuPRL2014, SchafferNatCommun2014,SignolesNaturePhysics2014,DebierrePRA2015,AlexanderPRA2015,QiuSciRep2015,SlichterNJP2016,MuellerSciRep2016,RuffoNJP2016, WuPRA2017,ZhouPRA2017,Chaudhryscirep2017a,Chaudhryscirep2017b,HanggiNJP2018,WuAnnals2018,Chaudhryscirep2018}.  On the other hand, the evolution of the quantum system can also be sped up if the measurements are not very frequent - this is known as the Quantum anti-Zeno effect (QAZE)~\cite{KurizkiNature2000, KoshinoPhysRep2005,BennettPRB2010,BaronePRL2004,YamamotoPRA2010,RaizenPRL2001}. To date, the majority of studies on QZE and QAZE have focused on population decay of quantum systems i.e. a system is prepared in an excited state and then is repeatedly measured to check if it is still in that state (see, for example, Refs.~\cite{KurizkiNature2000,KoshinoPhysRep2005,BennettPRB2010,BaronePRL2004,YamamotoPRA2010,RaizenPRL2001,ManiscalcoPRL2006,SegalPRA2007,ZhengPRL2008,AiPRA2010,ThilagamJMP2010,ThilagamJCP2013}). In this case, it is known that the decay rate of the quantum system depends on the overlap between the spectral density function of the environment and a measurement induced level width (see ref.~\cite{KurizkiNature2000}). This overlap changes with the measurement rate. A decrease in the measurement rate results in a crossover from the Zeno regime with large measurement rates to the anti-Zeno regime with smaller measurement rates. 

Moving beyond population decay models, we know that open quantum systems also undergo dephasing. In this context, the QZE and QAZE were studied for the exactly solvable pure dephasing model and significant differences were found with the population decay case \cite{ChaudhryPRA2014zeno}. This work was then generalized to developing a formalism that allows one to calculate the decay rate for arbitrary system-environment models \cite{Chaudhryscirep2016}. In this formalism, the decay rate is given by the overlap of the environment's spectral density and an effective `filter function' that depends on the system-environment model, the system state being repeatedly prepared, and the measurement interval. However, a major limitation is that the formalism assumes weak system-environment interaction. The study of the quantum Zeno and anti-Zeno effects for strong system-environment is relatively very much unexplored, although some progress has been made in this regard \cite{WuPRA2017,ZhouPRA2017,Chaudhryscirep2017a}. In particular, the QZE and QAZE have been studied in the spin-boson model with strong system-environment coupling strength in Ref.~\cite{Chaudhryscirep2017a}. The strategy used is to unitarily transform the Hamiltonian using a so called polaron transformation \cite{SilbeyJCP1984,VorrathPRL2005,ChinPRL2011,LeeJCP2012,LeePRE2012,GuzikJPCL2015} so that the system-environment interaction, in the new frame, is scaled by a system parameter such as the tunneling amplitude. Then, provided that the tunneling amplitude is sufficiently small, the effective system-environment interaction in the 'polaron' frame is weak and the problem becomes tractable in the polaron frame. The results in the strong coupling regime turn out to be qualitatively different from those in the weak coupling regime. For instance, it has been found that the decay rate in the strong coupling regime could no longer be written as an overlap of the spectral density and some filter function. Rather, the decay rate had a non-linear dependence on the spectral density \cite{Chaudhryscirep2017a}. A consequence of this is that an increase in the system-environment coupling strength results in a decrease in the decay rate, unlike the case of weak coupling regime. 

In this paper, we present a modified approach towards tackling the problem of investigating the QZE and QAZE in a spin-boson model with strong system-environment interaction. This approach is more powerful than the previous approaches since, in the weak coupling limit, it becomes equivalent to the weak coupling approach presented in Ref.~\cite{Chaudhryscirep2016}. On the other hand, our approach also becomes equivalent to the strong coupling approach of Ref.~\cite{Chaudhryscirep2017a} in the strong coupling limit. Furthermore, we can then use the approach we have presented to explore other regimes such as when the system-environment coupling is neither too large nor too small or when the value of the tunneling amplitude is not small. The strategy we have employed is again to perform a unitary transformation on the Hamiltonian and look at the physics in the the 'polaron' frame. However, the modification we make is that, following Ref.~\cite{NazirNJP2010} we now also keep a contribution of the system parameter $\Delta$ in the non-interacting part of the Hamiltonian in the new frame. In this way, if the environment is super-Ohmic, part of the contribution of the tunneling amplitude can be treated exactly, while the rest can be calculated perturbatively, thereby greatly expanding the regimes over which perturbation theory can be applied.

\section{Results and Discussion}

We start from a single spin 1/2 system coupled to an environment of harmonic oscillators. The system-environment Hamiltonian is (we set $\hbar = 1$ throughout) \cite{LeggettRMP1987,Weissbook,BPbook}
\begin{equation}
H_{L} = \frac{\varepsilon}{2} \sigma_{z} +  \frac{\Delta}{2}\sigma_{x} + \sum_{k} \omega_{k} {b^\dagger_{k}} b_{k} + \frac{\sigma_{z}}{2} \sum_{k} \biggl({g^*_{k}} {b_{k}} + g_{k} {b^\dagger_{k}} \biggr).
\notag
\end{equation}
Here $\varepsilon$ is the energy difference between the states of the two level system and $\Delta$ is the tunneling amplitude. The Hamiltonian can be broken down into the system part $H_{SL}=\frac{\varepsilon}{2} \sigma_{z}+\frac{\Delta}{2}\sigma_{x}$, the environment part $H_{BL}=\sum_{k} \omega_{k} {b^\dagger_{k}} b_{k}$, and the interaction part $H_{IL}=\frac{\sigma_{z}}{2} \sum_{k} ({g^*_{k}} {b_{k}} + g_{k} {b^\dagger_{k}})$.
If the system-environment coupling is weak, one can try to treat the interaction $H_{IL}$ perturbatively. However, since we aim to access the strong system-environment coupling regime as well, we do not follow this approach. Instead, we aim to unitarily transform the Hamiltonian such that in the new frame, the system-environment interaction becomes effectively weak. To be more concrete, the Hamiltonian can be unitarily transformed to the polaron frame \cite{Chaudhryscirep2017a}, so that the tunneling amplitude $\Delta$ appears in the new interaction term. We calculate $H = e^{\chi\sigma_{z}/2}H_{L}e^{-\chi\sigma_{z}/2}$, with $\chi = \sum_{k} (\alpha_{k} {b^\dagger_{k}} - {\alpha^*_{k}} {b_{k}} )$ and $\alpha_{k} = g_{k}/\omega_{k}$, to find that \cite{ SilbeyJCP1984,VorrathPRL2005,JangJCP2008,ChinPRL2011,LeeJCP2012,LeePRE2012,JangJCP2013,GuzikJPCL2015} 
\begin{equation}
H = \frac{\varepsilon}{2} \sigma_{z} + \sum_{k} \omega_{k} {b^\dagger_{k}} b_{k} + \frac{\Delta}{2} \biggl(\sigma_{+} B_{+} + \sigma_{-} B_{-}\biggr).
\label{oldpolaronH}
\end{equation}
In this way we can deal with strong coupling at the expense of treating $\Delta$ perturbatively. We call this the small delta approach. However, such an approach is questionable for larger values of $\Delta$. In this paper, our objective is to present an approach that allows us to access both the weak and strong system-environment coupling regimes even if the tunneling amplitude is not small. We are able to do this by performing the polaron transformation in a different manner such that the tunneling amplitude is present in both the system Hamiltonian and the system-environment interaction \cite{NazirNJP2010}. In this way, in addition to treating $\Delta$ perturbatively, there will be another term involving $\Delta$ that can be treated exactly. We can instead write Eq.~\eqref{oldpolaronH} as  
\begin{equation}
H = \frac{\varepsilon}{2} \sigma_{z} +  \frac{\Delta_{r}}{2}\sigma_{x} + \sum_{k} \omega_{k} {b^\dagger_{k}} b_{k} + \frac{\Delta}{2} \biggl(\sigma_{x} B_{x} + \sigma_{y} B_{y}\biggr),
\label{newpolaronH}
\end{equation}
where
\begin{flalign*}
B_{x} &= \frac{1}{2} (B_{+}+B_{-}-2B), && \\
B_{y} &= \frac{1}{2\dot{\iota}} (B_{-}-B_{+}),
\end{flalign*}
and
\begin{flalign*}
&B_{\pm} = \prod_{k} D(\pm\alpha_{k}), && \\
&D(\alpha_{k}) = e^{(\alpha_{k} {b^\dagger_{k}} - {\alpha^*_{k}} {b_{k}} )}.
\end{flalign*}
In these equations, $B=\langle B_{\pm} \rangle=\exp\left[-(1/2)\sum_k |\alpha_k|^2 \coth \left(\beta \omega_k/2\right) \right]$ and $\Delta_{r}=\Delta \times B$. It is straightforward to show that the Hamiltonian in eq.~\eqref{newpolaronH} and that in eq.~\eqref{oldpolaronH} are equivalent by plugging in the expansions of $B_x$ and $B_y$ in terms of $B_+$ and $B_-$. We now identify, in the polaron frame, $H_{S}=\frac{\varepsilon}{2} \sigma_{z} +  \frac{\Delta_{r}}{2}\sigma_{x}$, $H_{B}=\sum_{k} \omega_{k} {b^\dagger_{k}} b_{k}$ and $H_{I}=\frac{\Delta}{2} (\sigma_{x} B_{x} + \sigma_{y} B_{y})$. Clearly, the transformed Hamiltonian has $\Delta$ both in the `free' part $H_S$ and the interaction term. On the other hand, in Eq.~\eqref{oldpolaronH}, $\Delta$ is involved only in the interaction term. Our strategy now will be to treat $H_S$ exactly and $H_I$ perturbatively. Since part of the effect of the tunneling term is included now in $H_S$, we expect that we now need not be restricted to very small values of the tunneling amplitude when we apply perturbation theory. Our ultimate goal is to find the survival probability $S(N\tau)$, that is, the probability for the system to be measured in the state it started in after $N$ measurements with time interval $\tau$. Ignoring the build up of the system-environment correlations, we can write $S(t=N\tau)=\big(s(\tau)\big)^N$, where $s(\tau)$ is the survival probability after one measurement. This then allows us to define the effective decay rate $\Gamma(\tau)$ by the equation $S(t=N\tau)=e^{-\Gamma(\tau)  N  \tau}$, leading to 
\begin{equation}
\Gamma(\tau) = -\frac{1}{\tau} \ln\big(s(\tau)\big).
\label{decaygivensurvival}
\end{equation}
In order to compute $s(\tau)$, we need to write down the initial state of the system. The combined initial state of the system and the environment cannot be written as a product state because of strong system environment correlations \cite{ChaudhryPRA2013a,ChaudhryPRA2013b,Chaudhryscirep2017a}. The initial state of the system that we consider is $\ket{\uparrow}$, with $\sigma_z \ket{\uparrow} = \ket{\uparrow}$, which is the state usually used in the study of the population decay model \cite{KurizkiNature2000,KoshinoPhysRep2005,Chaudhryscirep2017a}. Since the system and the environment can be interacting strongly, the total state must be taken of the form $\rho_{L}(0)=P_{\uparrow} e^{- \beta H_{L}} P_{\uparrow}/Z$, where $P_\uparrow = \ket{\uparrow}\bra{\uparrow}$, and $Z=\text{Tr}_{S,B} \{P_{\uparrow} e^{- \beta H_{L} }\}$. On transformation to the polaron frame, if the system-environment interaction in the polaron frame is weak, the state reduces to a product state, $\rho(0)=\ket{\uparrow}\bra{\uparrow} \otimes e^{-\beta H_{B}}/Z_{B}$ where $Z_{B}=\text{Tr}_{B} \{ e^{- \beta H_{B} }\}$. To be more precise, $s(\tau)$ can be computed as
\begin{equation}
\label{probabilitylabframe}
s(\tau)= \text{Tr}_{S,B} \{\ket{\uparrow}\bra{\uparrow} e^{-i H_{L} \tau} \rho_{L}(0) e^{i H_{L} \tau}\}.
\end{equation}
On transforming to the polaron frame it becomes
\begin{equation}
\label{probabilitypolaronframe}
s(\tau)= \text{Tr}_{S,B} \{\ket{\uparrow}\bra{\uparrow} e^{-i H \tau} \rho(0) e^{i H \tau}\}.
\end{equation}
Or simply
\begin{equation}
s(\tau)= \text{Tr}_{S} \{\ket{\uparrow}\bra{\uparrow} \rho_{S}(\tau)\}=1-\bra{\downarrow}\rho_{S}(\tau)\ket{\downarrow}.
\label{survivalprob}
\end{equation}
From here onwards, we proceed using the perturbative approach.

We now sketch how to use perturbation theory to find the system density matrix at time $\tau$. We first break down the time evolution operator into two parts i.e. $U_{\text{tot}}(\tau)=e^{-\dot{\iota}H\tau}=U_0(\tau) U_I (\tau)$, where $U_0(\tau)$ is evolution due to the free part of the Hamiltonian and $U_I(\tau)$ is due to the interaction part. We can further separate $U_0(\tau)$ into the system part and the environment part i.e. $U_0(\tau)=U_S(\tau) U_B(\tau)$. Considering the interaction term as $\sum_\mu F_\mu \otimes B_\mu$, with $F_1 = \frac{\Delta}{2}\sigma_x$, $B_1 = B_x$, $F_2 = \frac{\Delta}{2}\sigma_y$ and $B_2 = B_y$, then using perturbation theory to second order gives $U_I(\tau) = \mathds{1} + A_1 + A_2$, where $A_1 = -\dot{\iota} \sum_\mu \int_0^\tau \widetilde{F}_\mu(t_1) \widetilde{B}_\mu(t_1) dt_1$ and $A_2 = -\sum_{\mu \nu} \int_0^\tau dt_1 \int_0^{t_1} dt_2 \widetilde{F}_\mu(t_1) \widetilde{F}_\nu(t_2) \widetilde{B}_\mu(t_1) \widetilde{B}_\nu(t_2)$. Here $\widetilde{F}_\mu(t) = U_S^\dagger(t) F_\mu U_S(t)$ and $\widetilde{B}_\mu(t)=U_B^\dagger(t) B_\mu U_B(t)$. Preserving terms only up to second order in $\Delta$, we get 
\begin{align}
\rho_S(\tau) \approx &\text{Tr}_B \lbrace U_0(\tau) [\rho(0) + \rho(0) A_1^\dagger + \rho(0)A_2^\dagger + \notag \\
&A_1 \rho(0) + A_2 \rho(0) + A_1 \rho(0) A_1^\dagger] U_0^\dagger(\tau) \rbrace.
\label{simplifythis}
\end{align}
As discussed before, we take $\rho(0) = \rho_S(0) \otimes \rho_B$. Simplifying the above equation term by term, we get 
\begin{align}
	\rho_S(\tau) = &U_S(\tau) \biggl( \rho_S(0) + \dot{\iota} \sum_\mu \int_0^\tau dt_1 [\rho_S(0),\widetilde{F}_\mu(t_1)]  \notag \\ 
	& \times \langle \widetilde{B}_\mu(t_1)\rangle_B + \sum_{\mu \nu}\int_0^\tau dt_1 \int_0^{t_1} dt_2 \bigl\lbrace C_{\mu \nu}(t_1, t_2) \notag \\ 
	& \times [\widetilde{F}_\nu(t_2)\rho_S(0),\widetilde{F}_\mu(t_1)] + \text{h.c.}\bigr\rbrace \biggr) U_S^\dagger (\tau).
\label{densitymatrixattau}
\end{align}
For future convenience, we write this as $\rho_S(\tau) = U_S(\tau) \rho_S^{\prime}(\tau) U_S^\dagger(\tau)$. Now, $U_S(\tau)$ can be written as
\begin{align*}
U_S(\tau) =& e^{-\dot{\iota} H_S \tau} = \mathds{1} \cos\Big(\frac{\Omega_{r} \tau}{2}\Big)- \\
& \dot{\iota} \Big(\frac{\Delta_{r}}{\Omega_{r}} \sigma_{x} +  \frac{\varepsilon}{\Omega_{r}} \sigma_{z} \Big) \sin\Big(\frac{\Omega_{r} \tau}{2}\Big).
\end{align*}
Putting this into eq.~\eqref{densitymatrixattau} and using eq.~\eqref{survivalprob}, the survival probability becomes
\begin{align}
s(\tau)& =  f(\tau) \: \bra{\downarrow} \rho_{S}^{\prime}(\tau) \ket{\downarrow} - g(\tau) \: \Big(\frac{\Delta_{r}}{\varepsilon}- \notag \\
& \text{Re}\left\{\bra{\downarrow} \rho_{S}^{\prime}(\tau) \ket{\uparrow}\right\}\Big) + h(\tau) \: \text{Im}\left\{\bra{\downarrow} \rho_{S}^{\prime}(\tau) \ket{\uparrow}\right\},
\label{simplifythistofindsurvival}
\end{align}
with
\begin{flalign}
&f(\tau) = \cos^{2}\Big(\frac{\Omega_{r} \tau}{2} \Big) + \frac{(\varepsilon^2 - \Delta_{r}^{2})}{\Omega_{r}^{2}} \sin^{2}\Big(\frac{\Omega_{r} \tau}{2} \Big)
\label{f} &&\\
&g(\tau) = -2\frac{\varepsilon \Delta_{r}}{\Omega_{r}^{2}} \sin^{2}\Big(\frac{\Omega_{r} \tau}{2} \Big) \label{g} \\
&h(\tau) = -\frac{\Delta_{r}}{\Omega_{r}} \sin(\Omega_{r} \tau ) .
\label{h}
\end{flalign}
Next, $\bra{\downarrow} \rho_{S}^{\prime}(\tau) \ket{\downarrow}$ and $\bra{\downarrow} \rho_{S}^{\prime}(\tau) \ket{\uparrow}$ have to be computed. To proceed we first calculate $\widetilde{F}_1(t)=\frac{\Delta}{2} \widetilde{\sigma}_x$ and $\widetilde{F}_2(t)=\frac{\Delta}{2} \widetilde{\sigma}_y$ as 
\begin{flalign*}
&\widetilde{\sigma}_{x}(t) = U_S^\dagger(t) \sigma_{x} U_S(t)
= a_{x}(t) \sigma_{x} + a_{y}(t) \sigma_{y} + a_{z}(t) \sigma_{z} &&\\
&\widetilde{\sigma}_{y}(t) = U_S^\dagger(t) \sigma_{y} U_S(t)
= b_{x}(t) \sigma_{x} + b_{y}(t) \sigma_{y} + b_{z}(t) \sigma_{z},
\end{flalign*}
where
\begin{flalign}
&a_{x}(t) = 1 - \frac{2 \varepsilon^2}{\Omega_{r}^{2}} \sin^{2}\Big(\frac{\Omega_{r} t}{2} \Big) \label{asx} &&\\
&a_{y}(t) = - \frac{\varepsilon}{\Omega_{r}} \sin(\Omega_{r} t) \label{asy} \\
&a_{z}(t) = \frac{2 \varepsilon \Delta_{r}}{\Omega_{r}^{2}} \sin^{2}\Big(\frac{\Omega_{r} t}{2} \Big) \label{asz}
\end{flalign}
\begin{flalign}
&b_{x}(t) = \frac{\varepsilon}{\Omega_{r}} \sin(\Omega_{r} t) \label{bsx} &&\\
&b_{y}(t) = \cos(\Omega_{r} t) \label{bsy} \\
&b_{z}(t) = - \frac{\Delta_{r}}{\Omega_{r}} \sin(\Omega_{r} t).
\label{bsz}
\end{flalign}
Moreover, the non-zero environment correlation functions are
\begin{flalign*}
&C_{11}(t) = \frac{B^2}{2} (e^{\phi(t)} + e^{-\phi(t)} - 2) &&\\
&C_{22}(t) = \frac{B^2}{2} (e^{\phi(t)} - e^{-\phi(t)}),
\end{flalign*}
where
\begin{flalign}
\phi(t) = \sum_k |\alpha_k|^2  \Big[&\cos(\omega_k t) \coth\Big(\frac{\beta \omega_k}{2}\Big) - \dot{\iota} \sin(\omega_k t)\Big].
\label{phi}
\end{flalign}
Given $\widetilde{F}_1(t)$, $\widetilde{F}_2(t)$ and the environment correlation functions, the rest of the calculation is straightforward though cumbersome. The full expression for the decay rate can be found in the appendix.

Before we compare the results of our approach with that of the previous approaches, let us comment on the approximate regime of validity of the approach we have taken above and to compare our approach with the previous weak and strong system-environment coupling approaches. First, we note that the effect of the environment on the system is captured fully by the spectral density function of the environment. The spectral density function contains within it the cutoff frequency $\omega_c$ that is related to the correlation time of the environment. Following Ref.~\cite{NazirNJP2010}, we expect that our approach is valid if 
\begin{align}
\Delta^4 \frac{|C_{\mu \nu}(0)|^2}{\omega_{c}^3} &\ll \Delta^2 \frac{|C_{\mu \nu}(0)|}{\omega_{c}},
\label{validitycondition}
\end{align} 
which implies
\begin{align}
\Big(\frac{\Delta}{\omega_{c}}\Big)^2 (1-B^4) &\ll 1.
\label{validitynew}
\end{align}
Both the strong and weak coupling regimes are captured. For the strong coupling regime, we expect that $B \rightarrow 0$ as the coupling strength is increased. This means that the condition above reduces to $\left(\frac{\Delta}{\omega_c}\right)^2 \ll 1$, which is what has been effectively considered in Ref.~\cite{Chaudhryscirep2017a}. On the other hand, in the weak coupling regime where the sytem-environment coupling strength is small, $B \rightarrow 1$, which again means that the above condition can still be satisfied. Furthermore, one can find system-environment parameters such that $\left(\frac{\Delta}{\omega_c}\right)^2 (1 - B^4)\ll 1$, while the applicability of the weak system-environment coupling approach or the small delta approach would be questionable.

\begin{figure}[h]
\begin{tabular}{@{}p{\linewidth}@{\quad}p{\linewidth}@{}}
    \subfigimg[width=\linewidth]{(a)}{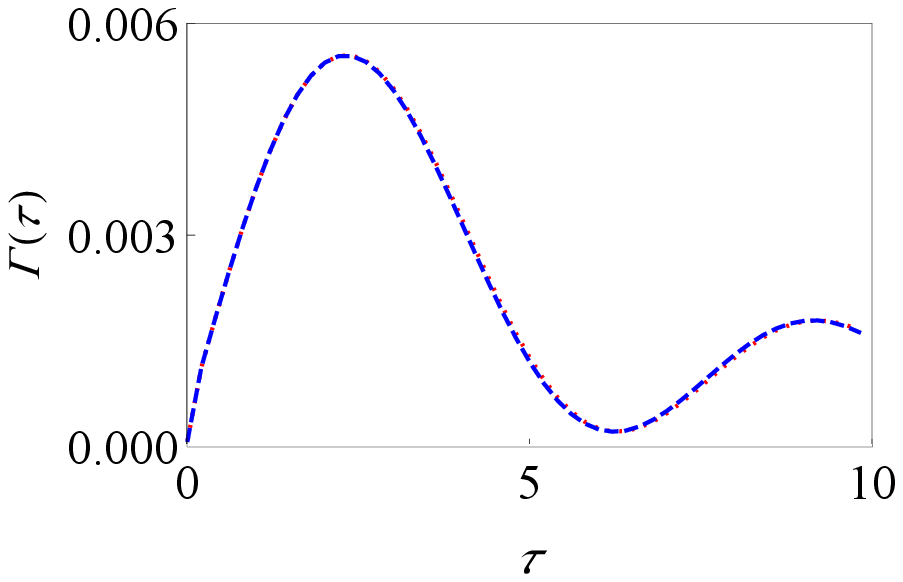} \\
    \subfigimg[width=\linewidth]{(b)}{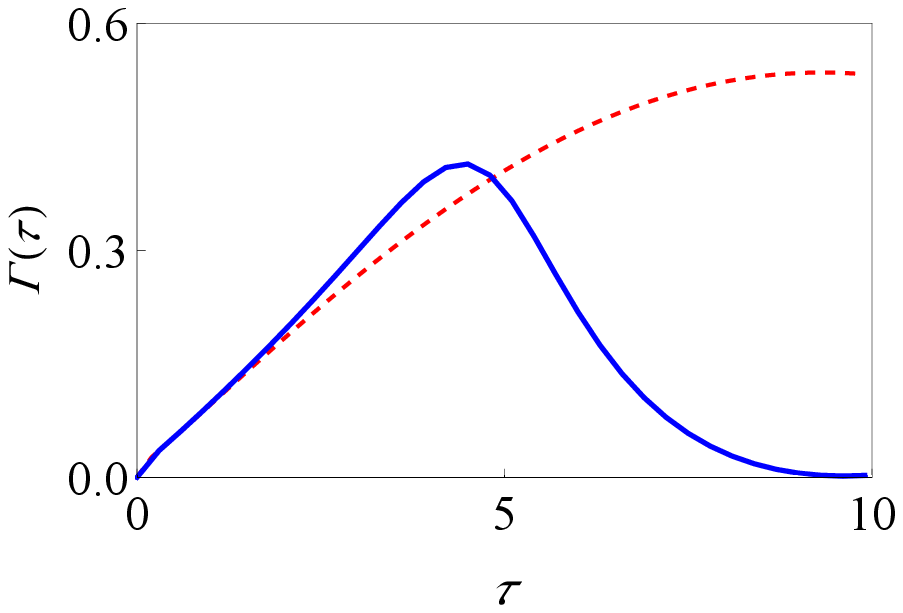} 
\end{tabular}
\caption{\textbf{Comparison with the small delta approach.} (a) $\Gamma$ as a function of $\tau$ for the small delta approach (red, dotted curve) and our approach (blue, dashed curve). Here we have used $G = 1$, $\varepsilon = 1$, $\omega_c = 10$, and $\Delta = 0.2$. The two curves have perfect agreement. (b) $\Gamma$ as a function of $\tau$ for the small delta approach (red, dashed curve) and our approach (solid, blue curve). Here we have used $G = 1$, $\varepsilon = 0.25$, $\omega_c = 10$, and $\Delta = 1$.}
\label{scComparison}
\end{figure} 

\begin{figure}[h]
\includegraphics[width=\linewidth]{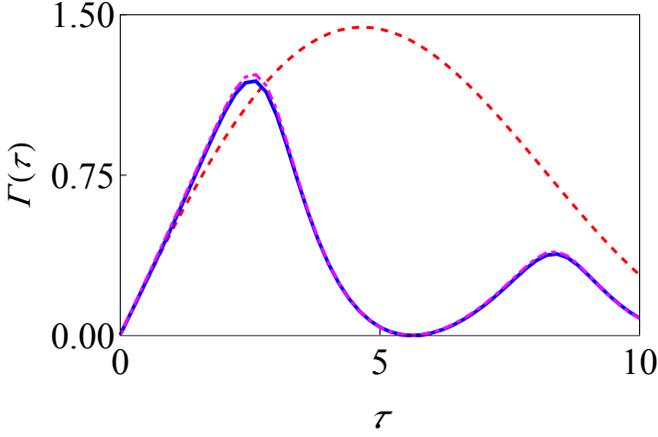}
\caption{\textbf{Comparison with the weak coupling approach.} Graph of $\Gamma$ as a function of $\tau$ with $G = 0.005$, $\varepsilon = 1$, $\omega_c = 10$, and $\Delta = 2$ for the small delta approach (dashed, red curve), the weak coupling approach (dot-dashed, magenta curve), and our approach (solid, blue curve). The blue and magenta curves lie on top of each other.}
\label{wcComparison}
\end{figure}

We now present numerical results. Let us first note that we can deduce Zeno and anti-Zeno regimes by looking at the behavior of $\Gamma(\tau)$ with $\tau$ (see ref.~\cite{Chaudhryscirep2016}). One possible approach is to compare $\Gamma(\tau)$ with the decay rate of the system when no measurement is performed on it. An alternative approach is to monitor the change in $\Gamma(\tau)$ with $\tau$. We say we are in the Zeno regime if $\Gamma(\tau)$ decreases by decreasing $\tau$ and in the anti-Zeno regime if $\Gamma(\tau)$ increases by decreasing $\tau$ (see, for example, Refs.~\cite{SegalPRA2007, ThilagamJMP2010, ChaudhryPRA2014zeno}). In this paper, we adopt the latter approach. Furthermore, the spectral density function of the environment $J(\omega)$ is now explicitly specified so that the sums over the environment modes can be evaluated via $\sum|g_k|^2 (\hdots) \rightarrow \int_0^\infty d\omega J(\omega) (\hdots)$. We consider $J(\omega) = G\omega^s \omega_c^{1-s} e^{-\omega/\omega_c}$ where $s$ is the Ohmicity of the environment (see ref.~\cite{BPbook}). $s=1$ corresponds to Ohmic, $s<1$ to sub-Ohmic and $s>1$ to a super-Ohmic environment respectively. It is important to note that for for Ohmic and sub-Ohmic environments, we find that $B=0$, which means $\Delta_{r}=0$. Consequently, for sub-Ohmic and Ohmic environment, our approach is exactly the same as the small delta approach - our approach only differs with super-Ohmic environments. The numerical results that we present consider zero temperature and $s = 3$. 

Fig.~\ref{scComparison} compares the results of our approach and the small delta approach in the strong-coupling regime. When the value of $\Delta$ is small compared to $\varepsilon$ [see Fig.~\ref{scComparison}(a)], both the approaches agree with each other exactly, showing that our approach reproduces the results of the small delta approach. As expected, if $\Delta$ is made larger [see Fig.~\ref{scComparison}(b)], the two approaches differ. More precisely, our approach predicts extra QZE-QAZE crossovers compared to the small delta approach. Let us now consider the opposite weak system-environment coupling regime. We show in Fig.~\ref{wcComparison} that in the weak coupling regime, our approach reproduces the results of the simple perturbation theory approach. The small delta approach (dashed, red curve) diverges from the weak coupling approach (dot-dashed, magenta curve) for $\Delta=2$, but our approach (solid, blue curve) still overlaps with the weak coupling approach exactly for all values of $\tau$. These results support our claim that the approach we have presented is more powerful than the previous two approaches because it can reproduce the results of the previous approaches in their respective regimes.

\begin{figure}[h]
\includegraphics[width=\linewidth]{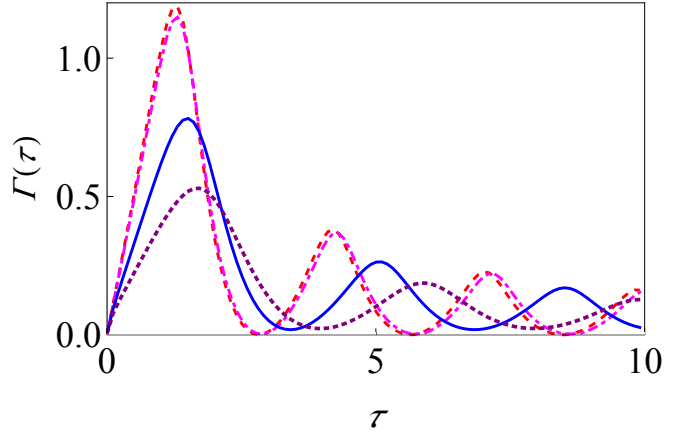}
\caption{\textbf{Variation of the decay rate with change in system-environment coupling strength.} Graph of $\Gamma$ as a function of $\tau$ with $\varepsilon = 1$, $\omega_c = 10$, and $\Delta = 2$ for $G = 0.01$ (dashed, red curve), $G = 0.05$ (dot-dashed, magenta curve), $G = 0.5$ (solid, blue curve) and $G = 0.95$ (dotted, purple curve) using our approach.}
\label{couplingComparison}
\end{figure}

After the benchmarking, we now use our approach to investigate the quantum Zeno and anti-Zeno effects in the moderate system-environment regime. Namely, the system-environment parameters are such that neither $G$ nor $\Delta$ are small, but it is still true that $(\Delta/\omega_c)^2(1 - B^4) \ll 1$. In Fig.~\ref{couplingComparison}, we show the variation of the decay rate as the system-environment coupling strength is varied. As we increase the coupling strength from $G = 0.5$ to $G = 0.95$, we observe that the decay rate decreases, similar to what has been noted in Ref.~\cite{Chaudhryscirep2017a}; however, we are now no longer restricted to small values of the tunneling amplitude $\Delta$ only. This decrease in the decay rate is non-intuitive because one would expect that as the system-environment coupling strength is increased, the decay rate increases. To understand this difference, we note that the expression for the survival probability depends on the system-environment Hamiltonian as well as the initial system-environment state [see Eq.~\eqref{probabilitylabframe}]. As the system-environment coupling strength is increased, the system-environment Hamiltonian $H_L$ obviously changes, but so does the initial state $\rho_L(0)$ as the system and the environment become more and more correlated. Note that we are taking these correlations into account - the initial system-environment product state in the polaron frame is a correlated system-environment state in the `lab' frame. As such, due to the correlations, the effective decay rate can actually decrease as the coupling strength increase; the system-environment correlations can actually protect the quantum state. Furthermore, as shown in Fig.~\ref{couplingComparison}, if $G$ is small, changing the system-environment coupling strength hardly changes the effective decay rate. The reason for this is simple. For weak system-environment coupling, the transition probability for the system is largely due to the system Hamiltonian itself, meaning that the transition probability, and hence the effective decay rate, is essentially the same. Consequently, we now analyze the decay of the initial system state in a different manner. If we are interested in the evolution due to the system-environment interaction, we should, before every measurement, remove the evolution due to the system Hamiltonian \cite{MatsuzakiPRB2010,ChaudhryPRA2014zeno,Chaudhryscirep2016,Chaudhryscirep2017a}. We can adopt a similar strategy even if the system-environment interaction is not weak. With this adjustment, the survival probability becomes
\begin{align*}
    s(\tau)=& \text{Tr}_{S,B} \{\ket{\uparrow}\bra{\uparrow} e^{i H_{SL} \tau} e^{-i H_L \tau} \rho_L(0) \\
    &e^{i H_L \tau} e^{-i H_{SL} \tau}\}.
\end{align*}
When we transform to the polaron frame, $H_{SL}$ is transformed into $H_S+H_I$. So, in the polaron frame,
\begin{align}
    s(\tau) = &\text{Tr}_{S,B} \{\ket{\uparrow}\bra{\uparrow} e^{i (H_{S}+H_I) \tau} e^{-i H \tau} \rho(0) \notag \\
    &e^{i H \tau} e^{-i (H_{S}+H_I) \tau}\}.
\label{survivalpolaronHsrmovaleq}
\end{align}

We have to solve Eq.~\eqref{survivalpolaronHsrmovaleq} to compute the survival probability. Since it has two different unitary evolution operators involving $H_I$, we'll have to write perturbation expansions for each of them separately. Again we'll have $U_{\text{tot}}(\tau) = e^{-\dot{\iota}H\tau} = U_0(\tau) U_I (\tau)$. With $F_1 = \frac{\Delta}{2}\sigma_x$, $B_1 = B_x$, $F_2 = \frac{\Delta}{2}\sigma_y$ and $B_2 = B_y$, the perturbation expansion to second order is $U_I (\tau) = \mathds{1} + A_1 + A_2$ where $A_1 = -\dot{\iota} \sum_\mu \int_0^\tau \widetilde{F}_\mu(t_1) \widetilde{B}_\mu(t_1) dt_1$ and $A_2 = -\sum_{\mu \nu} \int_0^\tau dt_1 \int_0^{t_1} dt_2 \widetilde{F}_\mu(t_1) \widetilde{F}_\nu(t_2) \widetilde{B}_\mu(t_1) \widetilde{B}_\nu(t_2)$ with $\widetilde{F}_\mu(t)=U_S^\dagger(t) F_\mu U_S(t)$ and $\widetilde{B}_\mu(t)=U_B^\dagger(t) B_\mu U_B(t)$. Similarly,  $e^{-\dot{\iota} (H_S+H_I) \tau} = U_S(\tau) U_I^{SP}(\tau)$. Then, $U_I^{SP}=\mathds{1} + A_1^{SP} + A_2^{SP}$ where $A_1^{SP} = -\dot{\iota} \sum_\mu \int_0^\tau \widetilde{F}_\mu(t_1) {B}_\mu dt_1$ and $A_2^{SP} = -\sum_{\mu \nu} \int_0^\tau dt_1 \int_0^{t_1} dt_2 \widetilde{F}_\mu(t_1) \widetilde{F}_\nu(t_2) {B}_\mu {B}_\nu$. So,
\begin{align*}
    &e^{i (H_{S}+H_I) \tau} e^{-i H \tau} \rho(0) e^{i H \tau} e^{-i (H_{S}+H_I) \tau} \\
    =& [\mathds{1} + (A_1^{SP})^\dagger + (A_2^{SP})^\dagger] U_B(\tau) [\rho_{\text{tot}}(0) + \rho(0) A_1^\dagger \\
    &+ \rho_{\text{tot}}(0)A_2^\dagger + A_1 \rho_{\text{tot}}(0) + A_2 \rho_{\text{tot}}(0) + \\
    &A_1 \rho(0) A_1^\dagger] U_B^\dagger(\tau) [\mathds{1} + A_1^{SP} + A_2^{SP}].
\end{align*}
When we keep terms to second order in $\Delta$, the only ones that survive are $U_B(\tau)A_1\rho(0)A_1^\dagger U_B^\dagger (\tau)$, \\$(A_1^{SP})^\dagger U_B(\tau) \rho(0)A_1^\dagger U_B^\dagger (\tau)$, $U_B(\tau)A_1\rho(0) U_B^\dagger (\tau) A_1^{SP}$ and $(A_1^{SP})^\dagger U_B(\tau)\rho(0) U_B^\dagger (\tau) A_1^{SP}$. The rest of the calculation proceeds easily. Again, the expression for decay rate comes out to be tedious and is provided in Appendix B.

\begin{figure}[h]
\begin{tabular}{@{}p{\linewidth}@{\quad}p{\linewidth}@{}}
    \subfigimg[width=\linewidth]{(a)}{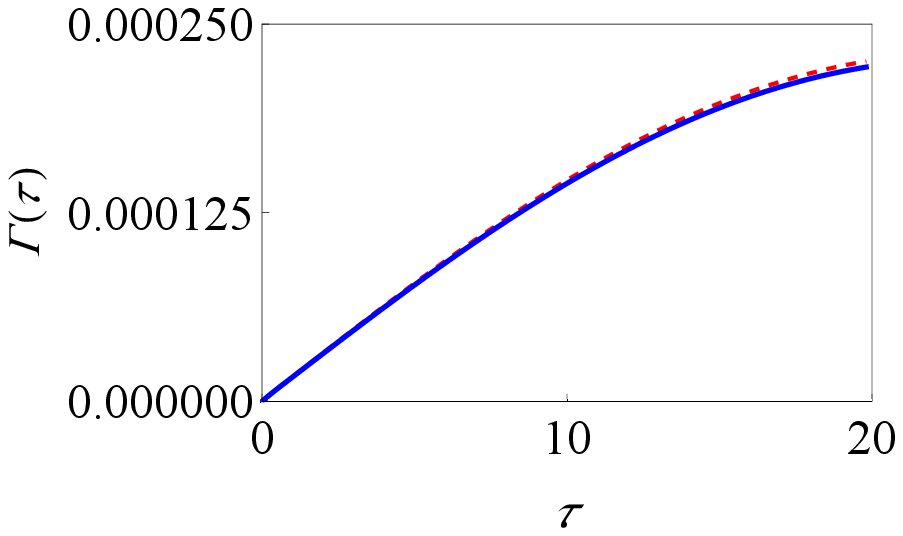} \\
    \subfigimg[width=\linewidth]{(b)}{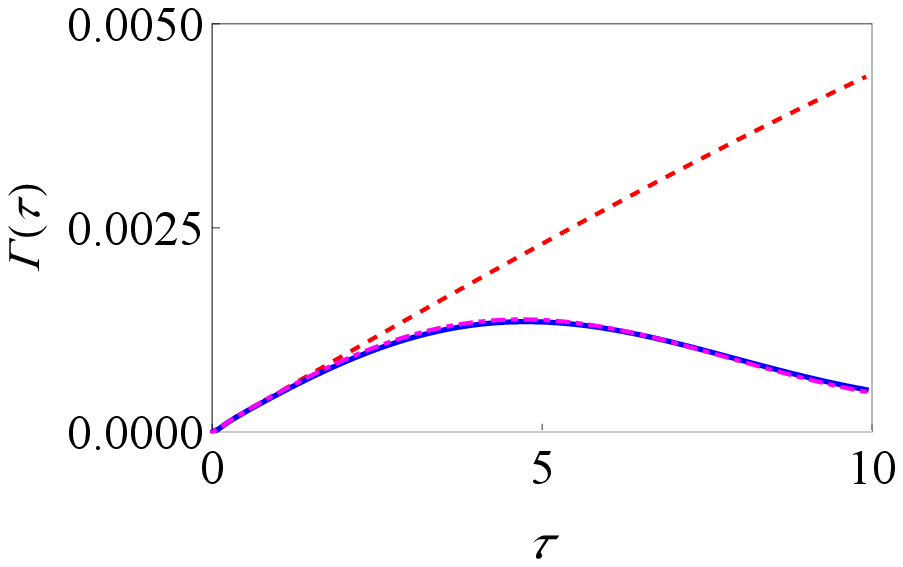}
\end{tabular}
\caption{\textbf{Comparison with the weak coupling and the small delta approaches after removing evolution due to system Hamiltonian.} (a) Graph of $\Gamma$ as a function of $\tau$ for the small delta approach (red, dashed curve) and our approach (solid, blue curve). Here we have used $G = 1$, $\varepsilon = 0.1$, $\omega_c = 10$, and $\Delta = 0.01$. Both the curves overlap exactly. (b) $\Gamma$ as a function of $\tau$ for the small delta approach (red, dashed curve), the weak coupling approach (dot-dashed, magenta curve) and our approach (solid, blue curve). Here we have used $G = 0.03$, $\varepsilon = 0.1$, $\omega_c = 10$, and $\Delta = 0.25$. The blue and magenta curves have an exact agreement.}
\label{HSRComparison}
\end{figure}

\begin{figure}[h!]
\begin{tabular}{@{}p{\linewidth}@{\quad}p{\linewidth}@{}}
    \subfigimg[width=\linewidth]{(a)}{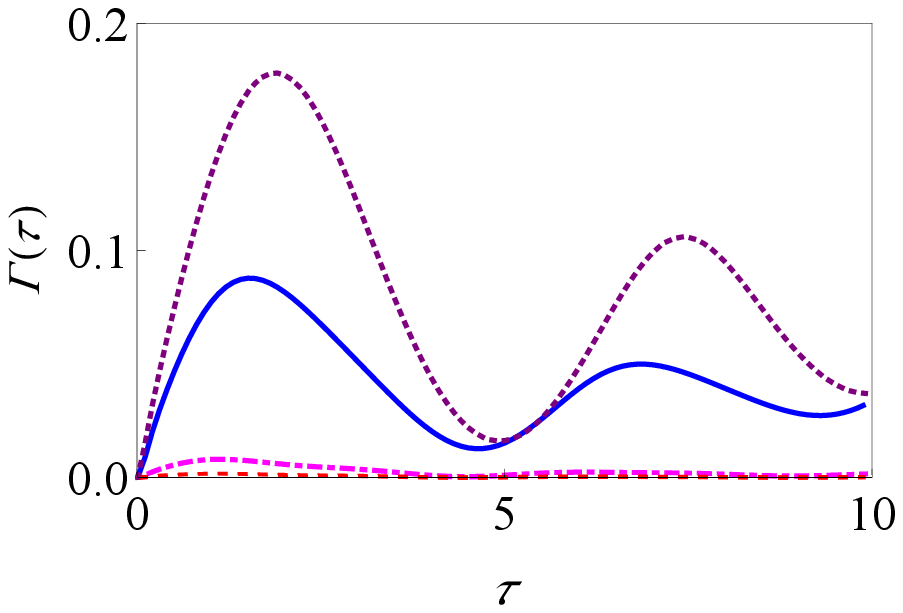} \\
    \subfigimg[width=\linewidth]{(b)}{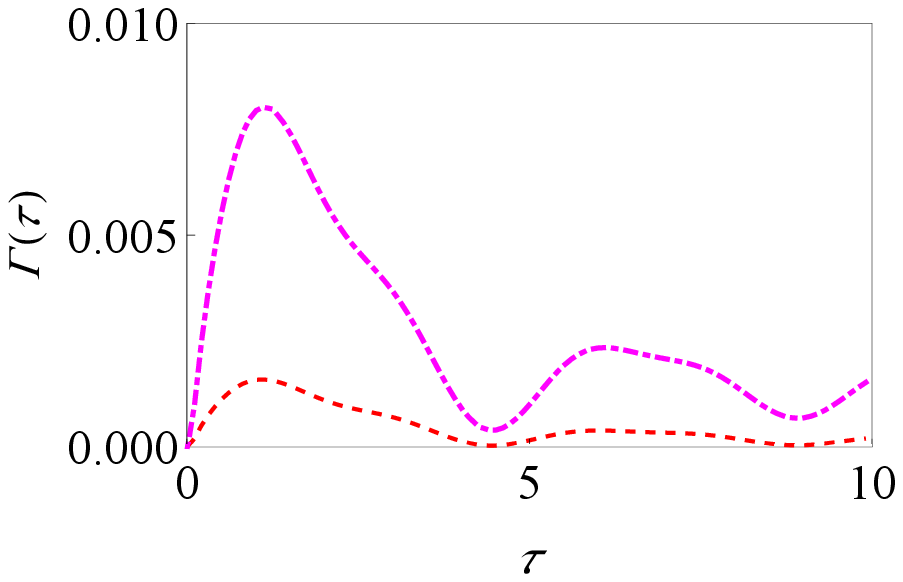} \\
    \subfigimg[width=\linewidth]{(c)}{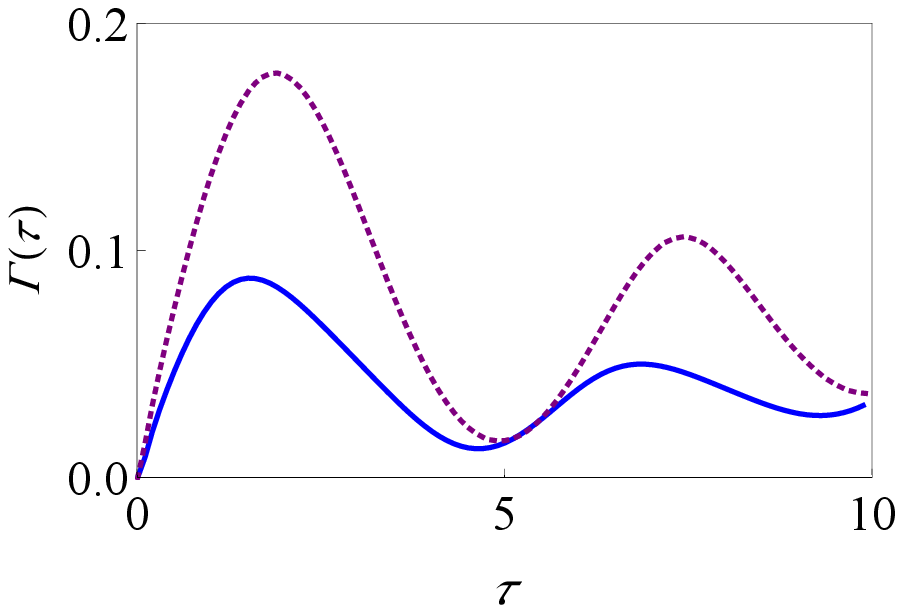}
\end{tabular}
\caption{\textbf{Variation of the decay rate with change in system-environment coupling strength after removing the evolution due to system Hamiltonian.} (a) Plots in (b) and (c) shown in one graph. (b) $\Gamma$ as a function of $\tau$ for $G = 0.01$ (dashed, red curve) and $G = 0.05$ (dot-dashed, magenta curve). We have used $\varepsilon = 1$, $\omega_c = 10$, and $\Delta = 1$. (c) Behaviour of $\Gamma$ for $G = 0.5$ (blue, solid curve) and $G = 0.95$ (dotted, purple curve) for the same parameters used in (b).}
\label{HSRcouplingchangeComparison}
\end{figure}

\begin{figure}[h]
\includegraphics[width=\linewidth]{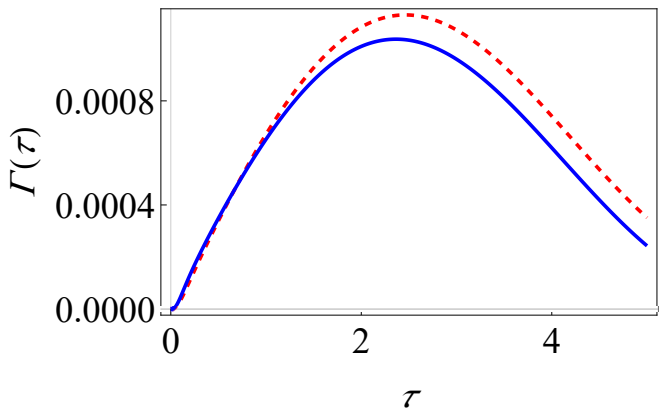}
\caption{\textbf{Variation of the decay rate with change in system-environment coupling strength for an Ohmic environment in the strong system-environment coupling regime.} Graph of $\Gamma$ as a function of $\tau$ with $\varepsilon = 1$, $\omega_c = 10$, and $\Delta = 0.05$ for $G = 1.5$ (dashed, red curve) and $G = 2$ (solid, blue curve).}
\label{ohmicfig}
\end{figure}

In Fig.~\ref{HSRComparison}(a), we have compared the results of our approach with the small delta approach in the strong coupling regime. Clearly, the results of the two approaches match perfectly. This is in accordance with our claim that for small values of $\Delta$ in the strong coupling regime, the two approaches become equivalent. Similarly, the results in the weak coupling regime have been compared in Fig.~\ref{HSRComparison}(b). In this case, even though the small delta approach (red, dashed curve) does not match with the weak coupling approach (magenta, dot-dashed curve), our approach (blue, solid curve) agrees with it as shown by the total overlap of the blue and magenta curves. Moving on, in Fig.~\ref{HSRcouplingchangeComparison}, we have illustrated the effect of varying the system-environment coupling strength on the effective decay rate. It is clear from Fig.~\ref{HSRcouplingchangeComparison}(b) that in the weak system-environment coupling regime, increasing the system-environment coupling strength increases the effective decay rate. Moreover, if we consider stronger system-environment coupling strengths, as done in Fig.~\ref{HSRcouplingchangeComparison}(c), then increasing the system-environment coupling strength also increases the effective decay rate now, in contrast with what happens if we do not remove the system evolution. It is clear then that in this case, with the super-Ohmic environment that we are considering, the system-environment correlations are not able to protect the system state. This is not always the case. With an Ohmic environment and small tunneling amplitude, as illustrated in Fig.~\ref{ohmicfig}, increasing the system-environment coupling strength can again decrease the effective decay rate.

\section{Conclusion}

We have worked out a new approach to study the QZE and QAZE for the spin boson model. Our approach provides corrections on the previous approaches for super-Ohmic environments. We have unitarily transformed the Hamiltonian to the `polaron' frame, so that the tunneling amplitude $\Delta$ appears in both the free part and the interaction part of the transformed Hamiltonian. Thereafter, we have used perturbation theory to derive an expression for the decay rate. We have shown that our approach is more powerful than the previous approaches because in addition to reproducing all the results of the previous approaches, it can also help us explore regimes which are beyond the scope of the previous approaches such as the moderate coupling regime. We have also worked out the expression for the effective decay rate of the quantum state after removing the evolution due to system Hamiltonian. Interestingly, we have shown that if the system-environment coupling strength is weak, the effective decay rate increases as the system-environment coupling strength is increased, while the opposite can be true when the system-environment coupling strength is strong.  





\newpage

\onecolumngrid

\appendix

\section{The expression for decay rate}
To find the expression for survival probability $s(\tau)$, we solve eq.~\eqref{simplifythistofindsurvival} to obtain
\begin{align*}
    s(\tau)=&1- \frac{\Delta_{r}^{2}}{\varepsilon^2+\Delta_R^2}\sin^{2}\Big(\frac{\Omega_{r} \tau}{2}\Big)-\frac{\Delta_{r}^{2}}{4} \Bigg\{\int_0^\tau dt \int_0^t dt^\prime e^{\phi_R(t^\prime)}.\bigg[\cos\big(\phi_I(t^\prime)\big).\bigg(f(\tau).\Big(a_x(t).a_x(t-t^\prime)\\
    &\quad\quad+a_y(t).a_y(t-t^\prime)+b_x(t).b_x(t-t^\prime)+b_y(t).b_y(t-t^\prime)\Big)+g(\tau).\Big(a_z(t).a_x(t-t^\prime)+b_z(t).b_x(t-t^\prime)\Big)\\
    &\quad\quad+h(\tau).\Big(a_z(t).a_y(t-t^\prime)+b_z(t).b_y(t-t^\prime)\Big)\bigg) +\sin\big(\phi_I(t^\prime)\big).\bigg(f(\tau).\Big(a_x(t).a_y(t-t^\prime)\\
    &\quad\quad-a_y(t).a_x(t-t^\prime)+b_x(t).b_y(t-t^\prime)-b_y(t).b_y(t-t^\prime)\Big)+g(\tau).\Big(a_z(t).a_y(t-t^\prime)-a_y(t).a_z(t-t^\prime)\\
    &\quad\quad+b_z(t).b_y(t-t^\prime)-b_y(t).b_z(t-t^\prime)\Big)+h(\tau).\Big(a_x(t).a_z(t-t^\prime)-a_z(t).a_x(t-t^\prime)\\
    &\quad\quad+b_x(t).b_z(t-t^\prime)-b_z(t).b_x(t-t^\prime)\Big)\bigg) \bigg]\\
    &+e^{-\phi_R(t^\prime)}.\bigg[\cos\big(\phi_I(t^\prime)\big).\bigg(f(\tau).\Big(a_x(t).a_x(t-t^\prime)+a_y(t).a_y(t-t^\prime)-b_x(t).b_x(t-t^\prime)-b_y(t).b_y(t-t^\prime)\Big)\\
    &\quad\quad+g(\tau).\Big(a_z(t).a_x(t-t^\prime)-b_z(t).b_x(t-t^\prime)\Big)+h(\tau).\Big(a_z(t).a_y(t-t^\prime)-b_z(t).b_y(t-t^\prime)\Big)\bigg) \\
    &\quad\quad-\sin\big(\phi_I(t^\prime)\big).\bigg(f(\tau).\Big(a_x(t).a_y(t-t^\prime)-a_y(t).a_x(t-t^\prime)-b_x(t).b_y(t-t^\prime)+b_y(t).b_y(t-t^\prime)\Big)\\
    &\quad\quad +g(\tau).\Big(a_z(t).a_y(t-t^\prime)-a_y(t).a_z(t-t^\prime)-b_z(t).b_y(t-t^\prime)+b_y(t).b_z(t-t^\prime)\Big)\\
    &\quad\quad +h(\tau).\Big(a_x(t).a_z(t-t^\prime)-a_z(t).a_x(t-t^\prime)-b_x(t).b_z(t-t^\prime)+b_z(t).b_x(t-t^\prime)\Big)\bigg) \bigg]\\
    &-2.\bigg[f(\tau).\Big(a_x(t).a_x(t-t^\prime)+a_y(t).a_y(t-t^\prime)\Big)+g(\tau).\Big(a_z(t).a_x(t-t^\prime)\Big)+h(\tau).\Big(a_z(t).a_y(t-t^\prime)\Big)\bigg]
    \Bigg\},
\end{align*}
where $f(t)$, $g(t)$, $h(t)$ are given in eq.~\eqref{f}, \eqref{g} and \eqref{h}. $a_x(t)$, $a_y(t)$, $a_z(t)$ are given in eq.~\eqref{asx}, \eqref{asy} and \eqref{asz}. and $b_x(t)$, $b_y(t)$, $b_z(t)$ in eq.~\eqref{bsx}, \eqref{bsy} and \eqref{bsz}. Moreover, $\phi_R(t)=\int_0^\infty d\omega \frac{J(\omega)}{\omega^2}\cos(\omega t) \coth\Big(\frac{\beta \omega}{2}\Big)$ and $\phi_I(t)=\int_0^\infty d\omega \frac{J(\omega)}{\omega^2} \sin(\omega t)$ so that $\phi(t)=\phi_R(t)-\dot{\iota}\phi_I(t)$ from eq.~\eqref{phi}. With the knowledge of $s(\tau)$, the decay rate $\Gamma(\tau)$ can be found using eq.~\eqref{decaygivensurvival}.

\section{The expression for decay rate after removing evolution due to system Hamiltonian}
The survival probability $s(\tau)$ for this case is obtained by simplifying Eq.~\eqref{survivalpolaronHsrmovaleq},
\begin{align*}
    s(\tau)=&1-\frac{\Delta_r^2}{4} \text{Re}\Bigg\{\int_0^\tau dt \int_0^t dt^\prime \bigg[\Big(e^{\phi(t^\prime)}-e^{\phi(t-\tau)}+e^{\phi(0)}\Big).\Big(\big(a_x(t)+\dot{\iota}a_y(t)\big).\big(a_x(t-t^\prime)-\dot{\iota}a_y(t-t^\prime)\big)\\
    & \quad \quad +\big(b_x(t)+\dot{\iota}b_y(t)\big).\big(b_x(t-t^\prime)-\dot{\iota}b_y(t-t^\prime)\big)\Big) \bigg] \\
    &+\bigg[\Big(e^{-\phi(t^\prime)}-e^{-\phi(t-\tau)}+e^{-\phi(0)}\Big).\Big(\big(a_x(t)+\dot{\iota}a_y(t)\big).\big(a_x(t-t^\prime)-\dot{\iota}a_y(t-t^\prime)\big)\\
    & \quad \quad -\big(b_x(t)+\dot{\iota}b_y(t)\big).\big(b_x(t-t^\prime)-\dot{\iota}b_y(t-t^\prime)\big)\Big)\bigg]\\
    &+\bigg[\Big(e^{\phi(t-t^\prime-\tau)}\Big).\Big(\big(a_x(t)-\dot{\iota}a_y(t)\big).\big(a_x(t-t^\prime)+\dot{\iota}a_y(t-t^\prime)\big)+\big(b_x(t)-\dot{\iota}b_y(t)\big).\big(b_x(t-t^\prime)+\dot{\iota}b_y(t-t^\prime)\big)\Big)\bigg]\\
    &-\bigg[\Big(e^{-\phi(t-t^\prime-\tau)}\Big).\Big(\big(a_x(t)-\dot{\iota}a_y(t)\big).\big(a_x(t-t^\prime)+\dot{\iota}a_y(t-t^\prime)\big)-\big(b_x(t)-\dot{\iota}b_y(t)\big).\big(b_x(t-t^\prime)+\dot{\iota}b_y(t-t^\prime)\big)\Big)\bigg]\Bigg\}.
\end{align*}

\twocolumngrid

\end{document}